# Atomic retrodiction

Stephen M. Barnett[1], David T. Pegg[2], John Jeffers[1] and Ottavia Jedrkiewicz[3]

[1]Department of Physics and Applied Physics, University of Strathclyde, Glasgow G4 0NG, U.K.

[2]School of Science, Griffith University, Nathan, Brisbane 4111, Queensland, Australia.

[3]Department of Physics, University of Essex, Wivenhoe Park, Colchester CO4 3SQ, U.K.

**Abstract.** Measurement of a quantum system provides information concerning the state in which it was prepared. In this paper we show how the retrodictive formalism can be used to evaluate the probability associated with any one of a given set of preparation events. We illustrate our method by calculating the retrodictive density operator for a two-level atom coupled to the electromagnetic field.

# 1. Introduction

Quantum theory is primarily about prediction, that is calculating the probabilities associated with any future measurement outcomes given information about the preparation, the subsequent evolution of the system and the measurement device to be used. A simple example is the probability that an initially excited atom, prepared at preparation time $t_p$, will be found by a suitable detector in its ground state at some measurement time $t_m$. This problem can be solved straightforwardly in the usual formalism, which we shall refer to in this paper as the *predictive* formalism. In this formalism, we speak of preparing the system in some state that depends only on the outcome of the preparation event. This state is most generally represented by a density operator that subsequently changes with time. Associated with the measuring device is, in general terms, a probability operator measure (POM) (Helstrom 1976). Projection of the density matrix at the time of measurement onto an element of the POM determines the probability of the outcome associated with that element. Ideal von Neumann measurements of an observable have possible measurement outcomes given by its eigenvalues and the associated POM element is simply the projector onto the corresponding eigenvector. In this formalism the state represents our knowledge of the preparation event.

Sometimes, for example in communication situations, we only have knowledge of the outcome of a measurement event and wish to deduce from this information about the preparation event. It is possible to use Bayes' theorem (Box and Tiao 1973) together with the predictive formalism for this purpose, but it is more direct to use the less common *retrodictive* formalism of quantum mechanics (Aharonov et al 1964, Penfield 1966, Belinfante 1975, Pegg and Barnett 1999, Barnett et al 2000a,b). In this formalism a state is assigned solely on the basis of the outcome of the measurement. We will refer to such a state as a retrodictive state. Thus instead of speaking of preparing the system in a state, we speak of measuring the system to be in a state. The retrodictive state transforms as we go back in time to the preparation event, at which time projection onto an operator associated with the preparation device determines the probability that the system was prepared in the state associated with that operator.

In previous work we have used the retrodictive formalism to study some problems in quantum optics including photon antibunching and EPR correlated photons, optical state preparation and quantum communications (Pegg and Barnett 1999, Barnett and Pegg 1999, Barnett et al 2000a,b). In this paper we shall exploit the formalism to study some preparation probabilities in a two-level atom interacting with an electromagnetic field.

**2. Retrodiction**

*2.1.   Closed systems*

We refer to a quantum mechanical system as closed if the system is prepared, evolves unitarily without interaction with another system and is then measured. A simple example involves a spin-half particle which can be prepared in an "up" or "down" state with a Stern-Gerlach apparatus. Another such apparatus can give a measurement result that the atom is in either of the states $|\pm\rangle$ corresponding, respectively, to the angles $\theta$ and $\pi + \theta$ to the up axis with no change of state between preparation and measurement. In one experiment we may know the outcome of the preparation event was the up state. We can then use the usual predictive formalism of quantum mechanics to calculate the probability that the outcome of the measurement corresponds to the state $|+\rangle$. The result is $|\langle \text{up}|+\rangle|^2$, that is, $\cos^2 \theta/2$.

In a second experiment we may be told that the outcome of the measurement corresponded to the state $|+\rangle$ but have no knowledge of the preparation event other than that the up and down states had equal *a priori* probabilities of being prepared. We wish to calculate, on the basis of this knowledge, the *a posteriori* probability that the atom was prepared in the down state. For this we would use the retrodictive formalism and assign the system the retrodictive state $|+\rangle$ just prior to the measurement. Knowing that the state does not change between preparation and measurement we would assign the retrodictive state $|+\rangle$ to the atom just after the preparation event. The probability that the atom was prepared in the down state is then $|\langle +|\text{down}\rangle|^2$, that is $\sin^2 \theta/2$. It is customary to speak of the event associated with the projection as the collapse of the wavefunction. In the predictive formalism, this takes place at the time of the measurement but in the retrodictive formalism it coincides with the preparation event.

We can treat the above simple example more formally in terms of a probability operator measure. The essential features of a POM are reviewed in Appendix A. For the case above, the measurement POM has two elements, $|+\rangle\langle+|$ and $|-\rangle\langle-|$ each associated with a possible measurement outcome and which sum to the unit operator. The retrodictive density operator associated with a particular outcome is just the corresponding POM element, normalised if necessary to have unit trace (Barnett *et al* 2000a). The retrodictive density operator associated with the outcome corresponding to detection of state $|+\rangle$ at the time of measurement is therefore simply $|+\rangle\langle+|$. As there is no evolution in this case between measurement and preparation, this is also the retrodictive density operator $\hat{\rho}_\theta^{\text{retr}}(t_\text{p})$ at the preparation time $t_\text{p}$. The source is *unbiased* because the up and down states have equal probability of being produced. The preparation POM elements (see appendix A and Barnett *et al* 2000a) $\hat{\Xi}_p$ are thus $|\text{up}\rangle\langle\text{up}|$ and $|\text{down}\rangle\langle\text{down}|$. The probability $P(\text{up}\,|\,+)$ that the atom was prepared in the up state if detected in the state $|+\rangle$ is $\text{Tr}\!\left[\hat{\rho}_\theta^{\text{retr}}(t_\text{p})\hat{\Xi}_\text{up}\right]$.

Suppose we change the example slightly so that the preparation apparatus includes some type of filter, say, which reduces the *a priori* probability $P(\text{down})$ for the down state to below one-half. Then instead of describing the (now *biased*) preparation device by POM elements $\hat{\Xi}_p$ we use instead the operators $\hat{\Lambda}_p$. These can be summed to form the *a priori* preparation density operator:

$$\hat{\Lambda} = \sum_p \hat{\Lambda}_p \quad . \tag{2.1}$$

In general the terms $\hat{\Lambda}_p$ can be associated with mixed states, but in our example here there are two terms, $\hat{\Lambda}_{\text{up}} = P(\text{up})|\text{up}\rangle\langle\text{up}|$ and $\hat{\Lambda}_{\text{down}} = P(\text{down})|\text{down}\rangle\langle\text{down}|$. In the absence of any further information as to what state was actually prepared, the best description we could give of the prepared state of the atom in the predictive formalism would be to assign it the predictive density operator $\hat{\Lambda}_{\text{up}} + \hat{\Lambda}_{\text{down}}$ based on our *a priori* knowledge. If we do know which state was actually prepared, say the up state, then we would assign the predictive density operator $\hat{\Lambda}_{\text{up}} / \text{Tr}(\hat{\Lambda}_{\text{up}})$. If, on the other hand, we do not know which state was actually prepared but know that the atom is detected in the state $|+\rangle$ we can calculate retrodictively the probability that the atom was prepared in the up state by projecting the retrodictive density operator corresponding to this measurement outcome onto the appropriate term $\hat{\Lambda}_p$ and obtain, after normalisation,

$$P(\text{up} \,|\, +) = \frac{\text{Tr}\!\left[\hat{\rho}_\theta^{\text{retr}}(t_{\text{p}})\hat{\Lambda}_{\text{up}}\right]}{\sum_p \text{Tr}\!\left[\hat{\rho}_\theta^{\text{retr}}(t_{\text{p}})\hat{\Lambda}_p\right]} \tag{2.2}$$

A corresponding expression holds for the probability that the atom was prepared in the down state if detected in the state $|+\rangle$.

We can summarise the predictive and retrodictive quantum formalisms for closed systems as follows. In the predictive formalism $\hat{\Lambda}_p$, associated with outcome *p* of the preparation device, becomes upon normalisation the predictive density operator. This operator evolves forward in time until it is projected onto $\hat{\Pi}_m$ to determine the probability of measurement outcome *m*. In the retrodictive formalism $\hat{\Pi}_m$, associated with outcome *m* of the measuring device, becomes upon normalisation the retrodictive density operator. This operator evolves backwards in time until it is projected onto $\hat{\Lambda}_p$ to determine the probability of preparation outcome *p*.

### 2.2. *Open systems*

An open system is one in which the subsystem *S* under study interacts with another subsystem. Often this second subsystem is larger and is referred to as the environment *E*. Normally we have at least some statistical information about the initial state of the environment and can assign it a density matrix. In the predictive formalism we can calculate the probability of obtaining a particular result of a measurement by finding the unitarily-evolved predictive density-operator for the combined system at the time of measurement and projecting this onto the appropriate combined POM element. In the usual case, in which we do not make any measurement of the environment, we can write the combined measurement POM element as proportional to the product of the POM element $\hat{\Pi}_m$ associated with the outcome $m$ for $S$ and the unit operator $\hat{1}_E$ on the space of the environment states. An alternative predictive approach is to represent the subsystem $S$ by a reduced density operator at the time of preparation, which is the trace of the density operator of the complete system over the states of the environment. The subsequent non-unitary evolution of the reduced density matrix is determined by a master equation. This equation is constructed so as to produce the correct measurement probability when the evolved reduced density operator at the time of the measurement is projected onto $\hat{\Pi}_m$. Thus, although the complete system usually evolves into an entangled state we can, in the usual predictive formalism, represent the state of $S$ in the limited sense that from it we can calculate the probability for $S$ to be detected in any particular state, conditioned on no information being obtained about the state of the environment.

In the retrodictive formalism of quantum mechanics, the retrodictive density operator of the system being measured is, immediately prior to the measurement, simply the POM element associated with the measurement outcome with a suitable normalisation factor (Barnett *et al* 2000a). Thus if the measurement gives an outcome $m$ for $S$ but no information about the environment, we can find the retrodictive density operator at an earlier time $t_p$ as

$$\hat{\rho}^{\text{retr}}(t_p) = \frac{\hat{U}^\dagger \hat{1}_E \hat{\Pi}_m \hat{U}}{\text{Tr}_{ES}[\hat{1}_E \hat{\Pi}_m]} \qquad (2.3)$$

where $\hat{U}$ is the forward time shift operator from the time $t_p$ to time $t_m$, the time of the measurement.

The terms $\hat{\Lambda}_p$ of the preparation device density operator for the system comprising $S$ and $E$ can be written as $\hat{\Lambda}_{p,S}\hat{\Lambda}_{p,E}$. If the preparation device is set up such that in the predictive formalism the environment always has the same initial state $\hat{\rho}_E^{\text{pred}}(t_p)$, then $\hat{\Lambda}_p = \hat{\Lambda}_{p,S}\hat{\rho}_E^{\text{pred}}(t_p)$. The probability that the system was prepared at time $t_p$ in the state corresponding to $\hat{\Lambda}_{p,S}$ is thus given by

$$\text{Tr}_{ES}\left[\hat{\rho}^{\text{retr}}(t_p)\hat{\Lambda}_p\right] = \frac{\text{Tr}_{ES}\left[\hat{U}^\dagger \hat{1}_E \hat{\Pi}_m \hat{U} \hat{\Lambda}_{p,S} \hat{\rho}_E^{\text{pred}}(t_p)\right]}{\sum_p \text{Tr}_{ES}\left[\hat{U}^\dagger \hat{\Pi}_m \hat{U} \hat{\Lambda}_{p,S} \hat{\rho}_E^{\text{pred}}(t_p)\right]}$$

$$= \frac{\text{Tr}_S\left[\hat{\rho}_S^{\text{retr}} \hat{\Lambda}_{p,S}\right]}{\sum_p \text{Tr}_S\left[\hat{\rho}_S^{\text{retr}} \hat{\Lambda}_{p,S}\right]} \qquad (2.4)$$

where

$$\hat{\rho}_S^{\text{retr}} = \frac{\text{Tr}_E\left[\hat{\rho}_E^{\text{pred}}(t_p) \hat{U}^\dagger \hat{\Pi}_m \hat{U}\right]}{\text{Tr}_{ES}[\hat{\rho}_E^{\text{pred}}(t_p) \hat{U}^\dagger \hat{\Pi}_m \hat{U}]} \quad . \qquad (2.5)$$

The reduced *retrodictive* density operator $\hat{\rho}_S^{\text{retr}}$ is of course not the same as the reduced *predictive* density operator. Nevertheless it is "reduced" in the sense that it acts only on the state space of *S* and not on the environment states. As with the reduced predictive density operator, this operator can be used to calculate only probabilities conditioned on the measurement not providing any information about the environment.

The main aim of this paper is to calculate $\hat{\rho}_S^{\text{retr}}$, which we shall refer to simply as the *retrodictive density operator*, for some open atomic systems. We can obtain the retrodictive density matrix elements in terms of a complete orthonormal basis set of states $\{|i\rangle\}$ of *S*. We have from (2.5)

$$\langle i|\hat{\rho}_S^{\text{retr}}|k\rangle = \frac{1}{N}\text{Tr}_E\left[\langle i|\hat{\rho}_E^{\text{pred}}(t_p)\hat{U}^\dagger\hat{\Pi}_m\hat{U}|k\rangle\right]$$

$$= \frac{1}{N}\text{Tr}_{ES}\left[\hat{\Pi}_m\hat{U}|k\rangle\langle i|\hat{\rho}_E^{\text{pred}}(t_p)\hat{U}^\dagger\right] \tag{2.6}$$

where we have used the cyclic property of the trace and *N* is the denominator in (2.5). By introducing a general time-independent operator in the form

$$\hat{A}_S = \sum_{i,k} A_{ik}|i\rangle\langle k|, \tag{2.7}$$

we can rewrite the retrodictive density matrix elements (2.6) as

$$\langle i|\hat{\rho}_S^{\text{retr}}|k\rangle = \frac{1}{N}\frac{\partial}{\partial A_{ki}}\text{Tr}_{ES}[\hat{\Pi}_m \hat{U}\hat{A}_S \hat{\rho}_E^{\text{pred}}(t_p)\hat{U}^\dagger]\ . \qquad (2.8)$$

If $\hat{A}_S$ has the form of an initial density matrix then we can find $\text{Tr}_{ES}[\hat{\Pi}_m \hat{U}\hat{A}_S \hat{\rho}_E^{\text{pred}}(t_p)\hat{U}^\dagger]$ from the appropriate master equation. We can most easily find the normalisation constant $N$ by noting that

$$\sum_i \langle i|\hat{\rho}_S^{\text{retr}}|i\rangle = \frac{1}{N}\sum_i \frac{\partial}{\partial A_{ii}}\text{Tr}_{ES}[\hat{\Pi}_m \hat{U}\hat{A}_S \hat{\rho}_E^{\text{pred}}(t_p)\hat{U}^\dagger] = 1. \qquad (2.9)$$

Expression (2.8) is useful for obtaining the retrodictive density matrix elements directly without first calculating preparation probabilities from (2.4). The retrodictive density matrix can then be used to calculate the preparation probability for any state by projection onto the associated preparation POM element $\hat{\Xi}_p$ or preparation density operator term $\hat{\Lambda}_p$ as appropriate.

An alternative approach, if we wish to calculate the preparation probability associated with a particular preparation POM element $\hat{\Xi}_p$ without explicitly finding the retrodictive density operator, is as follows. This probability is given by

$$\mathrm{Tr}_S\left(\hat{\rho}_S^{\mathrm{retr}}\hat{\Xi}_p\right) = \frac{1}{N}\mathrm{Tr}_{ES}\left[\hat{\rho}_E^{\mathrm{pred}}(t_p)\hat{U}^\dagger\hat{\Pi}_m\hat{U}\hat{\Xi}_p\right]. \tag{2.10}$$

We can rewrite this using the cyclic property of the trace as

$$\mathrm{Tr}_S\left(\hat{\rho}_S^{\mathrm{retr}}\hat{\Xi}_p\right) = \frac{1}{N}\mathrm{Tr}_S\left[\hat{\Pi}_m\hat{\Xi}_p(t_m)\right], \tag{2.11}$$

where

$$\hat{\Xi}_p(t_m) = \mathrm{Tr}_E\left[\hat{U}\hat{\Xi}_p(t_p)\hat{\rho}_E^{\mathrm{pred}}(t_p)\hat{U}^\dagger\right]. \tag{2.12}$$

If $\hat{\Xi}_p$ has the form of an initial density matrix we see from (2.12) that $\hat{\Xi}_p(t_m)$ has the form of a reduced density matrix in the predictive formalism obtainable from the appropriate master equation for the initial state of the environment (Barnett and Radmore 1997). It is straightforward to generalise this approach for a preparation density operator term $\hat{\Lambda}_p$ in place of $\hat{\Xi}_p$.

In general, finding the retrodictive density operator is more useful in that it allows the subsequent calculation of the preparation probability of any state and this is the approach we use for most of this paper. We do, however, illustrate the alternative approach of the above paragraph by finding directly the preparation state probabilities for a spontaneously emitting atom. For completeness, we also outline in Appendix B another method devised specifically for calculating the retrodictive density operator for a two-level atom.

## 3. Two-level atomic retrodiction

The retrodictive formalism for open systems allows us to calculate, from the outcome of a measurement, the probability associated with an earlier preparation event. In the preceding section we have presented formulae for obtaining directly either the probabilities for particular state preparations, or the retrodictive density matrix elements in any basis, which can then be used to calculate preparation probabilities. Here we shall illustrate our methods by using some examples from atomic physics, all based on the two-level atom coupled to the electromagnetic field. We work throughout in an interaction picture in which the energies of the two atomic levels do not appear in the Hamiltonian.

3.1. *Spontaneously emitting atom*

Consider first the simplest of open systems, the damped undriven two-level atom with excited and ground states $|e\rangle$ and $|g\rangle$. We assume that a measurement has been made

at time $t_m$ and that we know the state corresponding to the measurement result. We wish to retrodict a preparation event that has taken place at an earlier time $t_p = t_m - \tau$. Equation (2.11) gives the probability that the atom was prepared in a state corresponding to a particular preparation POM element. We simply need the appropriate measurement POM element and the precise form of $\hat{\Xi}_p(t_m)$. To be specific, suppose we find the atom to be in the ground state, so that $\hat{\Pi}_m = |g\rangle\langle g|$ is our measurement POM element. We seek, respectively, the probabilities that the atom was prepared in the excited state and the ground state, given that one of these two states was prepared at time $t_p$ by an unbiased apparatus so that the *a priori* probabilities for these two preparation events are 1/2. For a measurement made immediately after preparation ($\tau = 0$) the ground state preparation probability should be unity and the excited state preparation probability should vanish. In the other extreme, if there is a very long time between preparation and measurement, detecting the atom in the ground state should give us no information at all about the preparation state as all states decay eventually to the ground state. Thus the ground and excited state preparation probabilities given that we observe the atom in its ground state are 1/2 and the measurement has given us no information about the preparation of the atom. We now find the probability that the atom was prepared in the excited state at a general time $\tau$ prior to the measurement for which the corresponding preparation POM element is $\hat{\Xi}_p = |e\rangle\langle e|$. This element has the form of a density operator so we can find $\hat{\Xi}_p(t_\mathrm{m})$ from the usual predictive master equation (Barnett and Radmore 1997) obtained by taking the initial state of the environment $\hat{\rho}_\mathrm{E}^\mathrm{pred}(t_\mathrm{p})$ as the vacuum state. The solution to this master equation is

$$\hat{\Xi}_p(t_\mathrm{m}) = |e\rangle\langle e|\exp(-2\Gamma\tau) + |g\rangle\langle g|[1-\exp(-2\Gamma\tau)], \qquad (3.1)$$

where $\Gamma$ is one half of the Einstein A-coefficient for spontaneous emission. The excited state preparation probability conditioned on us detecting the atom in the ground state is then

$$P^\mathrm{prep}(e\,|\,g) = \frac{1}{N}\mathrm{Tr}_\mathrm{S}\left\{|g\rangle\langle g|[|e\rangle\langle e|\exp(-2\Gamma\tau) + |g\rangle\langle g|(1-\exp(-2\Gamma\tau))]\right\}$$

$$= \frac{1}{N}[1-\exp(-2\Gamma\tau)] \ . \qquad (3.2)$$

Similarly, the probability that the atom was prepared in the ground state can be calculated using $\hat{\Xi}_p = |g\rangle\langle g|$. This does not evolve with time as an undriven atom initially in the ground state must remain there. The ground state preparation probability thus becomes

$$P^\mathrm{prep}(g\,|\,g) = \frac{1}{N}\mathrm{Tr}_\mathrm{S}(|g\rangle\langle g|g\rangle\langle g|) = \frac{1}{N} \ . \qquad (3.3)$$

The probabilities (3.2) and (3.3) must sum to unity and hence

$$N = 2 - \exp(-2\Gamma\tau), \tag{3.4}$$

which gives

$$P^{\text{prep}}(e\,|\,g) = \frac{[1 - \exp(-2\Gamma\tau)]}{[2 - \exp(-2\Gamma\tau)]} \tag{3.5}$$

$$P^{\text{prep}}(g\,|\,g) = \frac{1}{[2 - \exp(-2\Gamma\tau)]} \ . \tag{3.6}$$

Clearly for $\tau = 0$, $P^{\text{prep}}(e\,|\,g) = 0$ and $P^{\text{prep}}(g\,|\,g) = 1$ as anticipated above. For $\tau \to \infty$, $P^{\text{prep}}(e\,|\,g) = P^{\text{prep}}(g\,|\,g) = 1/2$, so we have no preparation information. A similar calculation gives the excited and ground state preparation probabilities conditioned on us detecting the atom in the excited state. The results, $P^{\text{prep}}(e\,|\,e) = 1$ and $P^{\text{prep}}(g\,|\,e) = 0$ for any time interval between preparation and measurement, are intuitively obvious. If we detect the atom in the excited state then it cannot have decayed and it must, therefore, have been prepared in the excited state.

It is also possible to prepare and measure the atom in superpositions of the excited and ground states. Any measured superposition containing a non-zero ground state component gives no information about the prepared state if a long time has elapsed between preparation and measurement.

The above method for retrodicting probabilities required us to specify the preparation POM element in advance. A more versatile method is first to calculate the retrodictive density matrix associated with a measurement outcome. This density matrix can then be projected onto any preparation POM element. The retrodictive density matrix elements are readily obtained from equations (2.8) and (2.9). We write

$$\hat{A}_S = A_{ee}|e\rangle\langle e| + A_{gg}|g\rangle\langle g| + A_{eg}|e\rangle\langle g| + A_{ge}|g\rangle\langle e|, \tag{3.7}$$

and let this have the properties of a density operator. The appropriate master equation can then be solved to give

$$\begin{aligned}\hat{A}_S(t_m) &= \mathrm{Tr}_E\left[\hat{U}\hat{A}_S\hat{\rho}_E^{\mathrm{pred}}(t_p)\hat{U}^\dagger\right] \\ &= |e\rangle\langle e|A_{ee}\exp(-2\Gamma\tau) + |g\rangle\langle g|\{A_{gg} + A_{ee}[1-\exp(-2\Gamma\tau)]\} \\ &\quad + |e\rangle\langle g|A_{eg}\exp(-\Gamma\tau) + |g\rangle\langle e|A_{ge}\exp(-\Gamma\tau) \ .\end{aligned} \tag{3.8}$$

The retrodictive density matrix elements can now be found from (2.8). These elements are

$$\langle e|\hat{\rho}_S^{\text{retr}}|e\rangle = \frac{1}{N}\frac{\partial}{\partial A_{ee}}\text{Tr}_S\left[\hat{A}_S(t_m)\hat{\Pi}_m\right]$$

$$= \frac{1}{N}\left\{\exp(-2\Gamma\tau)\langle e|\hat{\Pi}_m|e\rangle + [1-\exp(-2\Gamma\tau)]\langle g|\hat{\Pi}_m|g\rangle\right\}, \quad (3.9)$$

$$\langle g|\hat{\rho}_S^{\text{retr}}|g\rangle = \frac{1}{N}\langle g|\hat{\Pi}_m|g\rangle, \quad (3.10)$$

$$\langle g|\hat{\rho}_S^{\text{retr}}|e\rangle = \frac{1}{N}\langle g|\hat{\Pi}_m|e\rangle\exp(-\Gamma\tau) \quad (3.11)$$

and

$$\langle e|\hat{\rho}_S^{\text{retr}}|g\rangle = \frac{1}{N}\langle e|\hat{\Pi}_m|g\rangle\exp(-\Gamma\tau) \quad . \quad (3.12)$$

The normalisation constant is such that the sum of the diagonal elements is unity. Setting the sum of (3.9) and (3.10) to unity gives

$$N = \exp(-2\Gamma\tau)\langle e|\hat{\Pi}_m|e\rangle + [2-\exp(-2\Gamma\tau)]\langle g|\hat{\Pi}_m|g\rangle \quad . \quad (3.13)$$

We note that for $\tau = 0$, the retrodictive density matrix is simply $\hat{\Pi}_m/\text{Tr}(\hat{\Pi}_m)$. This simply states that the retrodictive state immediately before the measurement is proportional to the POM element associated with the measurement outcome (Barnett et al. 2000a).

As an illustration, we use this retrodictive density matrix to calculate the probability $P^{\text{prep}}(e \mid g)$ that the excited state was prepared by an unbiased apparatus which is capable of preparing the atom in the excited or ground states, if the atom is detected to be in its ground state. Projecting the retrodictive density matrix onto the preparation POM element $|e\rangle\langle e|$ gives for this probability the expression (3.9) with $\hat{\Pi}_m = |g\rangle\langle g|$. This gives the result (3.5) obtained previously by direct calculation.

The retrodictive density matrix gives us more than just the ground and excited state probabilities. As an example, suppose that our measurement of the atom gave a result corresponding to the superposition state $|\theta\rangle = (\cos(\theta/2)|g\rangle + \sin(\theta/2)|e\rangle)$. Furthermore, let us suppose that we know that the atom was prepared in one of the non-orthogonal states $|e\rangle$ and $|+\rangle = (|e\rangle + |g\rangle)/\sqrt{2}$ and that the prior probabilities for these two states are $p$ and $1-p$ respectively. How does our measurement result change these probabilities? To answer this question we first calculate the retrodictive density matrix associated with our measurement result for which the POM element is

$$\hat{\Pi}_m = |\theta\rangle\langle\theta| = (\cos(\theta/2)|g\rangle + \sin(\theta/2)|e\rangle)(\cos(\theta/2)\langle g| + \sin(\theta/2)\langle e|). \qquad (3.14)$$

A little algebra then shows the retrodictive density matrix elements to be

$$\langle e|\hat{\rho}_S^{\text{retr}}|e\rangle = \frac{\frac{1}{2}\{1+\cos\theta[1-2\exp(-2\Gamma\tau)]\}}{1+\cos\theta[1-\exp(-2\Gamma\tau)]}, \qquad (3.15)$$

$$\langle g|\hat{\rho}_S^{\text{retr}}|g\rangle = \frac{\frac{1}{2}(1+\cos\theta)}{1+\cos\theta[1-\exp(-2\Gamma\tau)]} \qquad (3.16)$$

and

$$\langle g|\hat{\rho}_S^{\text{retr}}|e\rangle = \frac{\frac{1}{2}\sin\theta\exp(-\Gamma\tau)}{1+\cos\theta[1-\exp(-2\Gamma\tau)]} = \langle e|\hat{\rho}_S^{\text{retr}}|g\rangle. \qquad (3.17)$$

The two possible preparation events are associated with the operators

$$\hat{\Lambda}_e = p|e\rangle\langle e| \qquad (3.18)$$

and

$$\hat{\Lambda}_+ = (1-p)|+\rangle\langle +| = \frac{1}{2}(1-p)(|e\rangle+|g\rangle)(\langle e|+\langle g|). \qquad (3.19)$$

The probabilities that the atom was prepared in the state $|e\rangle$ or $|+\rangle$ given that it was subsequently observed to be in the state $|\theta\rangle$ are

$$P^{\text{prep}}(e\,|\,\theta) = \frac{\text{Tr}_S\left[\hat{\rho}_S^{\text{retr}}\hat{\Lambda}_e\right]}{\text{Tr}_S\left[\hat{\rho}_S^{\text{retr}}\left(\hat{\Lambda}_e+\hat{\Lambda}_+\right)\right]}$$

$$= \frac{p\{1+\cos\theta[1-2\exp(-2\Gamma\tau)]\}}{1+\cos\theta[1-\exp(-2\Gamma\tau)]+(1-p)\sin\theta\exp(-\Gamma\tau)-p\cos\theta\exp(-2\Gamma\tau)} \qquad (3.20)$$

and

$$P^{\text{prep}}(+|\theta) = \frac{\text{Tr}_S\left[\hat{\rho}_S^{\text{retr}}\hat{\Lambda}_+\right]}{\text{Tr}_S\left[\hat{\rho}_S^{\text{retr}}\left(\hat{\Lambda}_e + \hat{\Lambda}_+\right)\right]}$$

$$= \frac{(1-p)\{1+\cos\theta[1-\exp(-2\Gamma\tau)]+\sin\theta\exp(-2\Gamma\tau)\}}{1+\cos\theta[1-\exp(-2\Gamma\tau)]+(1-p)\sin\theta\exp(-\Gamma\tau)-p\cos\theta\exp(-2\Gamma\tau)}. \quad (3.21)$$

For $\tau = 0$, these probabilities are

$$P^{\text{prep}}(e|\theta) = \frac{p(1-\cos\theta)}{p(1-\cos\theta)+(1-p)(1+\sin\theta)}, \quad (3.22)$$

$$P^{\text{prep}}(+|\theta) = \frac{(1-p)(1+\sin\theta)}{p(1-\cos\theta)+(1-p)(1+\sin\theta)}. \quad (3.23)$$

These values agree with those obtained using Bayes' theorem directly (Barnett *et al* 2000a),

$$P^{\text{prep}}(e|\theta) = \frac{P(e,\theta)}{P(\theta)} = \frac{p|\langle e|\theta\rangle|^2}{p|\langle e|\theta\rangle|^2 + (1-p)|\langle +|\theta\rangle|^2} \quad (3.24)$$

$$P^{\text{prep}}(+|\theta) = \frac{P(+,\theta)}{P(\theta)} = \frac{(1-p)|\langle +|\theta\rangle|^2}{p|\langle e|\theta\rangle|^2 + (1-p)|\langle +|\theta\rangle|^2}. \quad (3.25)$$

For $\tau \to \infty$, the probabilities reduce to $P^{\text{prep}}(\text{e}|\theta) = p$ and $P^{\text{prep}}(+|\theta) = 1-p$, that is, the prior probabilities. This tells us that, after a long time, the measurement gives us no additional information about the initially prepared state.

3.2. *Incoherently excited atom*

We now consider the case of our two-level atom coupled to a thermal field of mean occupation number $\bar{n}$. This modifies the master equation governing the evolution of the density matrix (Barnett and Radmore 1997) and in place of (3.8) we obtain

$$\hat{A}_S(t_m) =$$

$$|e\rangle\langle e|\left\{A_{ee}\left[\frac{\bar{n}}{2\bar{n}+1} + \exp(-2\Gamma(2\bar{n}+1)\tau)\frac{\bar{n}+1}{2\bar{n}+1}\right] + A_{gg}\frac{\bar{n}}{2\bar{n}+1}[1-\exp(-2\Gamma(2\bar{n}+1)\tau)]\right\}$$

$$+|g\rangle\langle g|\left\{A_{gg}\left[\frac{\bar{n}+1}{2\bar{n}+1} + \frac{\bar{n}}{2\bar{n}+1}\exp(-2\Gamma(2\bar{n}+1)\tau)\right] + A_{ee}\frac{\bar{n}+1}{2\bar{n}+1}[1-\exp(-2\Gamma(2\bar{n}+1)\tau)]\right\}$$

$$+|e\rangle\langle g|A_{eg}\exp(-2\Gamma(2\bar{n}+1)\tau) + |g\rangle\langle e|A_{ge}\exp(-2\Gamma(2\bar{n}+1)\tau). \qquad (3.26)$$

The retrodictive density matrix elements can again be found by differentiation using (2.8). This procedure gives

$$\langle e|\hat{\rho}_S^{\text{retr}}|e\rangle = \frac{1}{N}\left\{\langle e|\hat{\Pi}_m|e\rangle\left[\frac{\bar{n}}{2\bar{n}+1} + \frac{\bar{n}+1}{2\bar{n}+1}\exp(-2\Gamma(2\bar{n}+1)\tau)\right]\right.$$

$$\left. +\langle g|\hat{\Pi}_m|g\rangle\frac{\bar{n}+1}{2\bar{n}+1}[1-\exp(-2\Gamma(2\bar{n}+1)\tau)]\right\}, \qquad (3.27)$$

$$\langle g|\hat{\rho}_S^{\text{retr}}|g\rangle = \frac{1}{N}\left\{\langle g|\hat{\Pi}_m|g\rangle\left[\frac{\bar{n}+1}{2\bar{n}+1} + \frac{\bar{n}}{2\bar{n}+1}\exp(-2\Gamma(2\bar{n}+1)\tau)\right]\right.$$

$$\left. +\langle e|\hat{\Pi}_m|e\rangle\frac{\bar{n}}{2\bar{n}+1}[1-\exp(-2\Gamma(2\bar{n}+1)\tau)]\right\}, \qquad (3.28)$$

$$\langle g|\hat{\rho}_S^{\text{retr}}|e\rangle = \frac{1}{N}\langle g|\hat{\Pi}_m|e\rangle\exp(-\Gamma(2\bar{n}+1)\tau) \qquad (3.29)$$

and

$$\langle e|\hat{\rho}_S^{\text{retr}}|g\rangle = \frac{1}{N}\langle e|\hat{\Pi}_m|g\rangle\exp(-\Gamma(2\bar{n}+1)\tau). \qquad (3.30)$$

The normalisation constant is

$$N = \langle e|\hat{\Pi}_m|e\rangle\left[\frac{2\bar{n}}{2\bar{n}+1} + \frac{1}{2\bar{n}+1}\exp(-2\Gamma(2\bar{n}+1)\tau)\right]$$

$$+\langle g|\hat{\Pi}_m|g\rangle\left[\frac{2\bar{n}+2}{2\bar{n}+1} - \frac{1}{2\bar{n}+1}\exp(-2\Gamma(2\bar{n}+1)\tau)\right]. \qquad (3.31)$$

For $\tau \to \infty$, it is clear that the retrodictive density operator tends to one half of the identity operator, corresponding to no information about the initially prepared state. This happens for all measured states because the thermal radiation can cause absorption so that it is possible to prepare the atom in its ground state and to find it later in its excited state. Figure 1 shows the predictive and retrodictive diagonal density matrix elements for an atom respectively prepared or measured to be in its excited state. The predictive density matrix elements rapidly decay to their steady state form for which the ratio of the excited and ground state probabilities is $\bar{n}/(1+\bar{n})$. The retrodictive diagonal density matrix elements tend to the value 1/2 for long time intervals $\tau$.

### 3.3. *Coherently driven atom*

In our final example the atom is resonantly driven by an intense laser field. It is well known that such a system exhibits Rabi oscillations at a frequency $\Omega = (V^2 - \Gamma^2/4)^{1/2}$ where $V$ is proportional to the laser field strength and $\Gamma$ is the atomic spontaneous decay rate. The (interaction picture) Hamiltonian describing the coupling between the atom and the laser field is

$$\hat{H} = \frac{V}{2}(\hat{\sigma}_+ + \hat{\sigma}_-), \qquad (3.32)$$

where $\hat{\sigma}_+ = |e\rangle\langle g|$ and $\hat{\sigma}_- = |g\rangle\langle e|$. The reduced, predictive density operator for the atom satisfies the master equation

$$\frac{\partial}{\partial t}\hat{\rho}_S^{pred} = -i\left[\frac{V}{2}(\hat{\sigma}_+ + \hat{\sigma}_-), \hat{\rho}_S^{pred}\right] + \Gamma\left[2\hat{\sigma}_-\hat{\rho}_S^{pred}\hat{\sigma}_+ - \hat{\sigma}_+\hat{\sigma}_-\hat{\rho}_S^{pred} - \hat{\rho}_S^{pred}\hat{\sigma}_+\hat{\sigma}_-\right]. \qquad (3.33)$$

The solution for the predictive reduced density operator at the measurement time may be written in the form

$$\hat{\rho}_S^{pred}(t_m) = \frac{1}{2}\left[\hat{1} + u(\tau)\hat{\sigma}_1 + v(\tau)\hat{\sigma}_2 + w(\tau)\hat{\sigma}_3\right], \qquad (3.34)$$

where $\hat{1} = |e\rangle\langle e| + |g\rangle\langle g|$ is the unit operator on the space of the atom. The Pauli operators $\hat{\sigma}_1 = |e\rangle\langle g| + |g\rangle\langle e|$, $\hat{\sigma}_2 = i(|g\rangle\langle e| - |e\rangle\langle g|)$ and $\hat{\sigma}_3 = |e\rangle\langle e| - |g\rangle\langle g|$ represent respectively the real and imaginary parts of the dipole moment and the atomic inversion. The functions $u(\tau)$, $v(\tau)$ and $w(\tau)$ are the expectation values of $\hat{\sigma}_1$, $\hat{\sigma}_2$ and $\hat{\sigma}_3$ respectively at time $\tau$ after preparation and have the values (Barnett and Radmore 1997)

$$u(\tau) = u(0)\exp(-\Gamma\tau) \qquad (3.35)$$

$$v(\tau) = \frac{1}{V^2 + 2\Gamma^2}\Big(2\Gamma V + \exp(-3\Gamma\tau/2)\{[(V^2 + 2\Gamma^2)v(0) - 2\Gamma V]\cos(\Omega\tau)$$
$$+ \Omega^{-1}[-3\Gamma^2 V + (V^2 + 2\Gamma^2)(-Vw(0) + \Gamma v(0)/2)]\sin\Omega\tau\}\Big) \qquad (3.36)$$

$$w(\tau) = \frac{1}{V^2 + 2\Gamma^2}\Big(-2\Gamma^2 + \exp(-3\Gamma\tau/2)\{[2\Gamma^2 + (V^2 + 2\Gamma^2)w(0)]\cos\Omega\tau$$

$$+\Omega^{-1}[-2\Gamma(V^2+\Gamma^2/2)+(V^2+2\Gamma^2)(Vv(0)-\Gamma w(0)/2)]\sin\Omega\tau\}\big). \qquad (3.37)$$

The excited and ground state matrix elements can be found from the master equation solution (3.34) using

$$\langle e|\hat{\rho}_S^{pred}(t_p)|e\rangle = A_{ee} = [1+w(0)]/2 \qquad (3.38)$$

$$\langle g|\hat{\rho}_S^{pred}(t_p)|g\rangle = A_{gg} = [1-w(0)]/2 \qquad (3.39)$$

$$\langle e|\hat{\rho}_S^{pred}(t_p)|g\rangle = A_{eg} = [u(0)-iv(0)]/2 \qquad (3.40)$$

$$\langle g|\hat{\rho}_S^{pred}(t_p)|e\rangle = A_{ge} = [u(0)+iv(0)]/2. \qquad (3.41)$$

These can be substituted into (3.34)-(3.37) which allows us to calculate the retrodictive density matrix elements according to (2.8). The calculations are somewhat lengthy to reproduce in detail, so we shall concentrate on the results for a few specific measurement POM elements.

We first study the case where the atom is detected to be in its excited state, so the measurement POM element is $\hat{\Pi}_m = |e\rangle\langle e|$. The retrodictive matrix elements are then

$$\langle e|\hat{\rho}_S^{\text{retr}}|e\rangle = \frac{1}{2N}\left(1 + \frac{\partial w(t)}{\partial A_{ee}}\right) = \frac{1}{2N(V^2 + 2\Gamma^2)}\{V^2 + \exp(-3\Gamma\tau/2)[(V^2 + 4\Gamma^2)\cos\Omega\tau$$

$$- \frac{\Gamma}{2\Omega}(5V^2 + 4\Gamma^2)\sin\Omega\tau]\}, \qquad (3.42)$$

$$\langle g|\hat{\rho}_S^{\text{retr}}|g\rangle = \frac{V^2}{2N(V^2 + 2\Gamma^2)}\left[1 - \exp(-3\Gamma\tau/2)\left(\cos(\Omega\tau) + \frac{3\Gamma}{2\Omega}\sin(\Omega\tau)\right)\right], \qquad (3.43)$$

$$\langle e|\hat{\rho}_S^{\text{retr}}|g\rangle = -\langle g|\hat{\rho}_S^{\text{retr}}|e\rangle = \frac{-iV}{2N\Omega}\exp(-3\Gamma\tau/2)\sin(\Omega\tau), \qquad (3.44)$$

where the normalisation constant is $1 + \frac{1}{2}\left(\frac{\partial w(\tau)}{\partial A_{ee}} + \frac{\partial w(\tau)}{\partial A_{gg}}\right)$. The evolution of the predictive and retrodictive matrix elements for this system is shown in figure 2. In the retrodictive case the diagonal matrix elements show Rabi oscillations which decay to the "no information" level of 1/2. The off-diagonal elements decay to zero as expected. This corresponds to a loss of coherence information about the system a long time in the past. Similar behaviour is found for $\hat{\Pi}_m = |g\rangle\langle g|$.

Suppose instead that the atom is measured in the superposition $\frac{1}{\sqrt{2}}(|e\rangle + i|g\rangle)$, which gives $\hat{\Pi}_m = \frac{1}{2}(1 + \hat{\sigma}_2)$. The retrodictive matrix elements are

$$\langle e|\hat{\rho}_S^{\text{retr}}|e\rangle = \frac{1}{2N}\left\{1 + \frac{1}{V^2 + 2\Gamma^2}\left[2\Gamma V\right.\right.$$

$$-\exp(-3\Gamma\tau/2)\left(2\Gamma V\cos(\Omega\tau)+\frac{V}{\Omega}(V^2+5\Gamma^2)\sin(\Omega\tau)\right)\right]\right\}, \tag{3.45}$$

$$\langle g|\hat{\rho}_S^{\text{retr}}|g\rangle = \frac{1}{2N}\left\{1+\frac{1}{V^2+2\Gamma^2}\left[2\Gamma V\right.\right.$$

$$\left.\left.-\exp(-3\Gamma\tau/2)\left(2\Gamma V\cos(\Omega\tau)+\frac{V}{\Omega}(\Gamma^2-V^2)\sin(\Omega\tau)\right)\right]\right\}, \tag{3.46}$$

$$\langle e|\hat{\rho}_S^{\text{retr}}|g\rangle = -\langle g|\hat{\rho}_S^{\text{retr}}|e\rangle = \frac{-i\exp(-3\Gamma\tau/2)}{2N}\left[\cos(\Omega\tau)+\frac{\Gamma}{2\Omega}\sin(\Omega\tau)\right],$$

(3.47)

with $N = 1 + \frac{1}{2}\left(\frac{\partial v(\tau)}{\partial A_{ee}}+\frac{\partial v(\tau)}{\partial A_{gg}}\right)$. These are shown in figure 3, along with the corresponding predictive evolution. At the measurement time the retrodictive diagonal matrix elements take their long-time values of 1/2. The off-diagonal matrix elements are pure imaginary, and so they undergo Rabi oscillations. These elements couple to the diagonal matrix elements so that we also see oscillations in the diagonal density matrix elements.

Suppose now that the atom is measured to be in the superposition $\frac{1}{\sqrt{2}}(|e\rangle+|g\rangle)$, so that the measurement POM element, $\hat{\Pi}_m = \frac{1}{2}(1+\hat{\sigma}_1)$, contains only the real part of the dipole matrix element. In this case the retrodictive matrix elements are much simpler,

$$\langle e|\hat{\rho}_S^{\text{retr}}|e\rangle = \langle g|\hat{\rho}_S^{\text{retr}}|g\rangle = \frac{1}{2}, \tag{3.48}$$

$$\langle e|\hat{\rho}_S^{\text{retr}}|g\rangle = \langle g|\hat{\rho}_S^{\text{retr}}|e\rangle = \frac{1}{2}\exp(-\Gamma\tau). \tag{3.49}$$

Here the diagonal matrix elements are time-independent. The reason is that the off-diagonal matrix elements are purely real and so do not lead to Rabi oscillations, that is $u(\tau)$ simply decays.

The predictive state of a driven two-level atom evolves forward in time towards a steady state with an unchanging population inversion and imaginary dipole matrix element. The expectation values of the Pauli operators $\hat{\sigma}_1$, $\hat{\sigma}_2$ and $\hat{\sigma}_3$ for this state are

$$u(\tau) = 0, \tag{3.50}$$

$$v(\tau) = \frac{2\Gamma V}{V^2 + 2\Gamma^2}, \tag{3.51}$$

$$w(\tau) = \frac{-2\Gamma^2}{V^2 + 2\Gamma^2}. \tag{3.52}$$

If we were to perform a measurement showing the atom to be in this state then we could calculate a corresponding retrodictive density operator. The diagonal and off-diagonal matrix elements of this density operator are shown in figures 4a and 4b respectively. The off-diagonal matrix elements have a non-zero imaginary part so the retrodictive density matrix exhibits damped Rabi oscillations. The long time interval limit of the retrodictive density operator is, once again, one half of the identity operator, corresponding to providing no information about the initially prepared state.

## 4. Conclusions

The retrodictive formalism provides the means to evaluate preparation probabilities given the results of some later measurement event. Our approach relies on the introduction of a retrodictive state. This is a state assigned to the system prior to the measurement, based on the measurement result. Immediately before the measurement, this retrodictive state is simply the normalised POM element corresponding to the measurement result (Barnett *et al* 2000a). Put more simply, the retrodictive density operator is the state corresponding to the measurement result. If the system is closed, that is isolated from its environment, then the change in time of the retrodictive density operator is unitary. If, however, the system is coupled to an unmonitored environment then we will lose information about the system through coupling to the environment. This will mean in most cases that the retrodictive density operator for long times prior to the measurement will tend to become proportional to the identity operator. This is the zero information state and reflects the fact that our measurement tells us nothing about the way the system was prepared long before the measurement.

We have applied the formalism to retrodict from measured states of a two-level atom interacting with the electromagnetic field. For an undriven atom, which can only undergo spontaneous emission, we find that the diagonal elements tend to the value 1/2 and the off-diagonal ones to zero. This reflects the fact that we can obtain from the measurement no information about the system a long time before the measurement. The only exception is if the atom is measured in its excited state. In this case, the retrodictive density operator must be the excited state projector $|e\rangle\langle e|$ for all times prior to the measurement. This is because there is no light to absorb and so the atom could not have made a transition to the excited state from the ground state. With this exception, the decay to "no information" is a general feature of retrodiction for damped two-level atoms, both driven and undriven.

We have also performed retrodictive calculations for the incoherently and coherently driven atom. The evolution of the retrodictive density matrices of both systems is different in character from that of their predictive counterparts. In the forward time direction an atom driven by an incoherent field relaxes to a steady state predictive density operator with the ratio of the excited and ground state probabilities of $\bar{n}/(\bar{n}+1)$. The corresponding ratio for retrodictive evolution is unity. For the coherently driven atom we find retrodictive Rabi oscillations which decay. Again the predictive and retrodictive evolutions decay to different forms for long time intervals, with the retrodictive state tending to the zero-information state with density operator $\hat{1}/2$.

**Acknowledgements**

We thank Prof. R. Loudon for helpful comments and for a careful reading of our manuscript. This work was supported by the United Kingdom Engineering and Physical Sciences Research Council and by the Australian Research Council.

**Appendix A. Probability operator measures**

A comprehensive discussion of a probability operator measure (POM) is given by Helstrom (1976). Here we outline only the essential features.

In the predictive formalism we can use a POM as a mathematical representation of a measuring device. The POM is a set of elements $\hat{\Pi}_m$, each of which is a positive semi-definite operator associated with a different possible outcome of a measurement. $\hat{\Pi}_m$ is defined such that the probability of the outcome $m$ from a measurement of a system with a predictive density matrix $\hat{\rho}^{\text{pred}}$ at the time of measurement is given by $\text{Tr}[\hat{\rho}^{\text{pred}}\hat{\Pi}_m]$. The total probability for all possible outcomes must be unity. Thus

$$\sum_m \text{Tr}[\hat{\rho}^{\text{pred}}\hat{\Pi}_m] = \text{Tr}[\hat{\rho}^{\text{pred}} \sum_m \hat{\Pi}_m]$$

$$= 1. \tag{A.1}$$

For this to be true for all $\hat{\rho}^{\text{pred}}$, normalisation of the trace of $\hat{\rho}^{\text{pred}}$ to unity requires $\sum_m \hat{\Pi}_m$ to equal the unit operator.

A measurement that yields no information may be considered as a measurement with only one possible outcome, for example a zero meter reading, for all $\hat{\rho}^{\text{pred}}$. Normalisation of the POM as above requires the corresponding single element to be the unit operator.

We can also use a POM with elements $\hat{\Xi}_p$ as a mathematical description of an unbiased preparation device. In this case $\text{Tr}[\hat{\rho}^{\text{retr}}\hat{\Xi}_p]$ is the *a posteriori* probability of a preparation outcome *p* based on knowledge of the measurement outcome. This probability is equal to that obtainable from the predictive formalism combined with Bayes' theorem (Barnett *et al* 2000a). If the preparation device is biased we work instead with components $\hat{\Lambda}_p$ of the *a priori* density matrix.

**Appendix B. Alternative derivation of retrodictive density matrix**

Here we outline an alternative method for finding the retrodictive density matrix for a two-level atom coupled to the environment. We first expand the retrodictive density operator for the atom in terms Pauli operators $\hat{\sigma}_j$ with $j = 1, 2, 3$ (Barnett and Radmore 1997). The general form of this expansion is

$$\hat{\rho}_S^{\text{retr}} = \frac{1}{2}\left(\hat{1} + \sum_j u_j \hat{\sigma}_j \right) \tag{B.1}$$

where $\hat{1}$ is the unit operator on the space of the atom. Multiplying by $(\hat{1} + \hat{\sigma}_k)$ and tracing over the atom states we obtain

$$\text{Tr}_S[(\hat{1} + \hat{\sigma}_k)\hat{\rho}_S^{\text{retr}}] = 1 + u_k . \tag{B.2}$$

Thus from (2.5) we have

$$u_k = \frac{1}{N}\text{Tr}_{ES}[(\hat{1} + \hat{\sigma}_k)\hat{\rho}_E^{\text{pred}}(t_p)\hat{U}^\dagger \hat{\Pi}_m \hat{U}] - 1$$

$$= \frac{1}{N}\text{Tr}_{ES}[\hat{U}(\hat{1} + \hat{\sigma}_k)\hat{\rho}_E^{\text{pred}}(t_p)\hat{U}^\dagger \hat{\Pi}_m] - 1$$

$$= \frac{1}{N}\text{Tr}_S[(\hat{1} + \hat{\sigma}_k)(t_m)\hat{\Pi}_m] - 1 \tag{B.3}$$

where, because $(\hat{1} + \hat{\sigma}_k)/2$ has the form of a density matrix, we can find

$$(\hat{1} + \hat{\sigma}_k)(t_\mathrm{m}) = \mathrm{Tr}_\mathrm{E}[\hat{U}(\hat{1} + \hat{\sigma}_k)\hat{\rho}_\mathrm{E}^\mathrm{pred}(t_\mathrm{p})\hat{U}^\dagger] \tag{B.4}$$

from the appropriate master equation. Indeed the values of $(\hat{1} + \hat{\sigma}_k)(t_\mathrm{m})$ can be obtained immediately from the general expressions given by Barnett and Radmore (1997), that is, equations (3.35)-(3.37) with values of $u(0)$, $v(0)$ and $w(0)$ chosen as 0 or 1 as appropriate. Inserting these into (B.1) gives the retrodictive density matrix in the form

$$\hat{\rho}_\mathrm{S}^\mathrm{retr} = \frac{1}{2}\left(\hat{1} + \sum_j \left\{\frac{1}{N}\mathrm{Tr}_\mathrm{S}[(\hat{1} + \hat{\sigma}_j)(t_\mathrm{m})\hat{\Pi}_m] - 1\right\}\hat{\sigma}_j\right). \tag{B.5}$$

Finally we note that the expression for $N$, which is given by the denominator of (2.5) can be written as

$$N = \mathrm{Tr}_\mathrm{ES}[\hat{U}\hat{\rho}_\mathrm{E}^\mathrm{pred}(t_\mathrm{p}) \otimes \hat{1}\hat{U}^\dagger \hat{\Pi}_m]$$

$$= \mathrm{Tr}_\mathrm{S}[\hat{1}(t_\mathrm{m})\hat{\Pi}_m] \tag{B.6}$$

where, because $\hat{1}/2$ has the form of a density operator, we can find $\hat{1}(t_\mathrm{m}) = \mathrm{Tr}_\mathrm{E}\left[\hat{U}\hat{\rho}_\mathrm{E}^\mathrm{pred}(t_\mathrm{p}) \otimes \hat{1}\hat{U}^\dagger\right]$ from the same master equation as used above.

**Figure Captions**

Fig. 1. Incoherently excited atom with a mean thermal occupation number of $\bar{n} = 1$. (a) Predictive diagonal density matrix elements plotted against time for an atom initially prepared in the excited state. (b) Retrodictive diagonal density matrix elements for an atom measured to be in the excited state. The solid and dashed curves are the excited and ground state matrix elements respectively. The time $\tau$, measured in units of the atomic state lifetime, is the interval between preparation and measurement. Note that in all figures, the preparation time is to the left of the measurement time. This means that retrodictive density matrix elements are plotted with $\tau$ increasing to the left.

Fig. 2. Coherently driven atom with $V = 4\Gamma$. (a) Predictive diagonal density matrix elements for an initially excited atom, and (b) retrodictive diagonal density matrix elements for an atom measured to be in the excited state. The solid and dashed curves are the excited and ground state matrix elements respectively. (c) Imaginary part of the predictive and (d) retrodictive off-diagonal matrix elements for the same systems as (a) and (b). The solid and dashed curves are the imaginary parts of $\langle e|\hat{\rho}_S^{\text{retr}}|g\rangle$ and $\langle g|\hat{\rho}_S^{\text{retr}}|e\rangle$ respectively. The real parts are zero.

Fig. 3. Coherently driven atom with $V = 4\Gamma$. (a) Predictive diagonal density matrix elements for an atom initially in the superposition $(|e\rangle + i|g\rangle)/\sqrt{2}$, and (b) retrodictive diagonal density matrix elements for an atom measured to be in the same superposition. (c) Imaginary part of the predictive and (d) retrodictive off-diagonal matrix elements for the same systems as (a) and (b). The solid and dashed curves represent the same quantities as in figure 2.

Fig. 4. Coherently driven atom with $V = 4\Gamma$. (a) Retrodictive diagonal density matrix elements for an atom measured to be in the predictive steady state. (b) Retrodictive off-diagonal matrix elements for the same system. The solid and dashed curves represent the same quantities as in figures 2b and 2d.

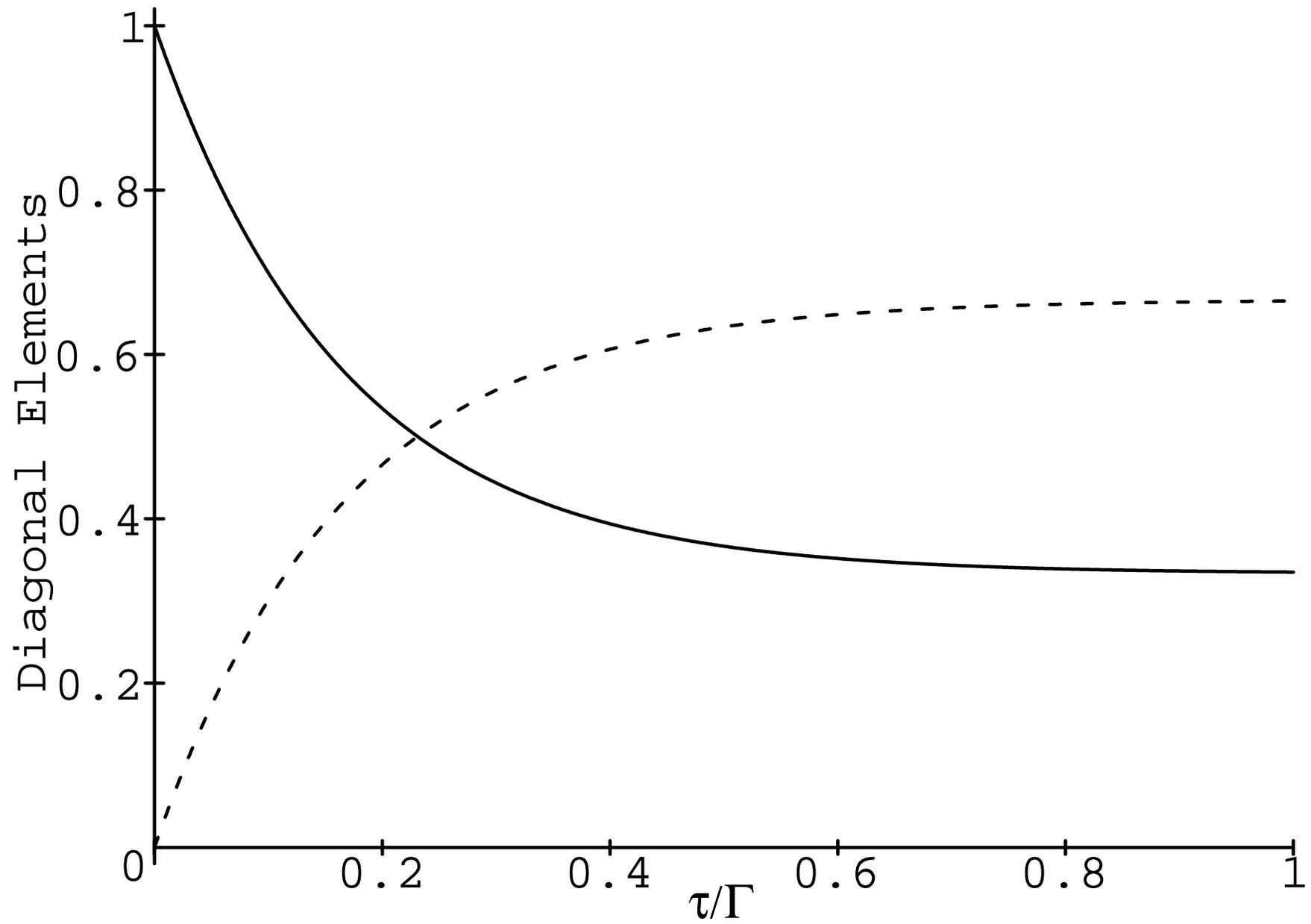

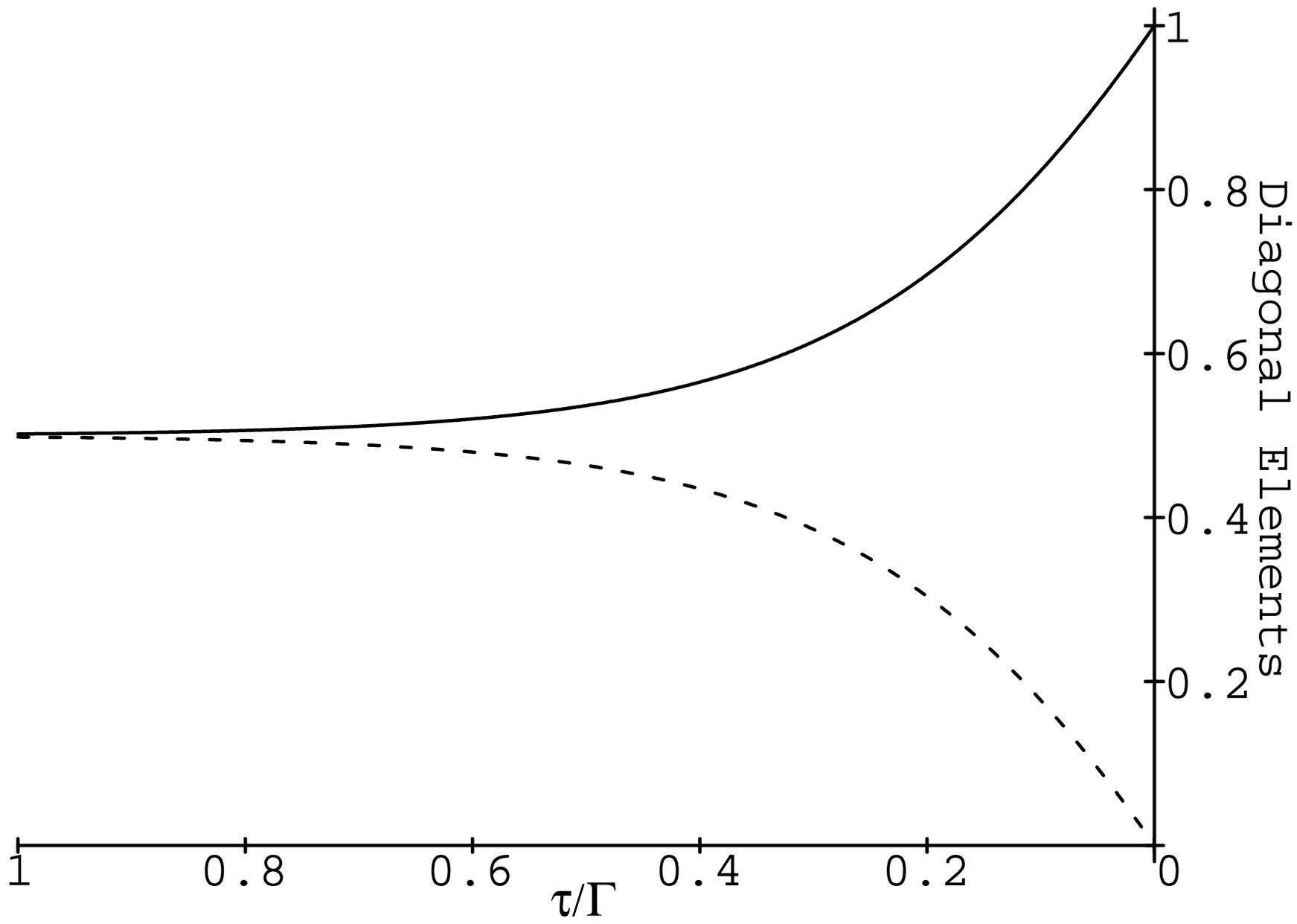

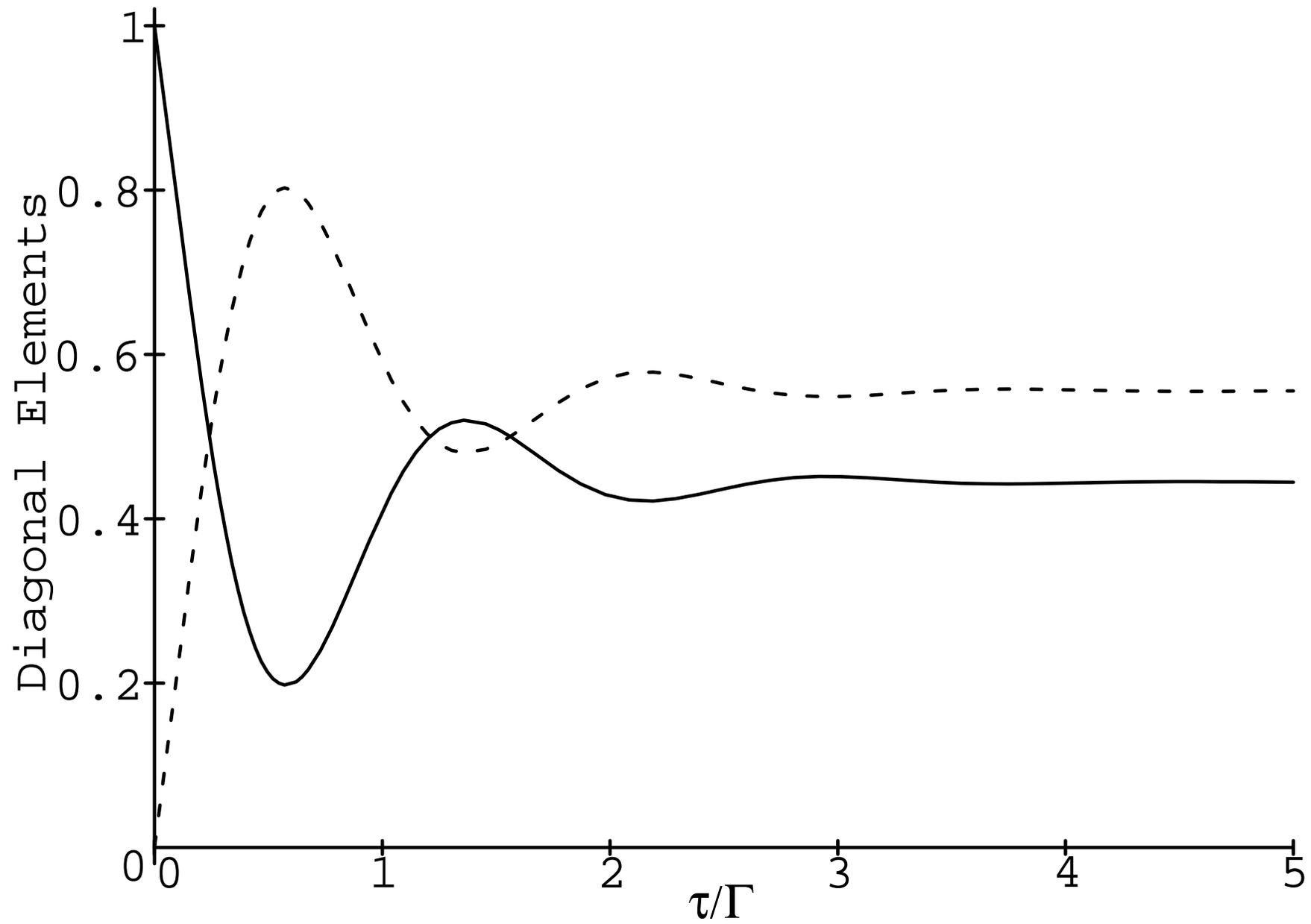

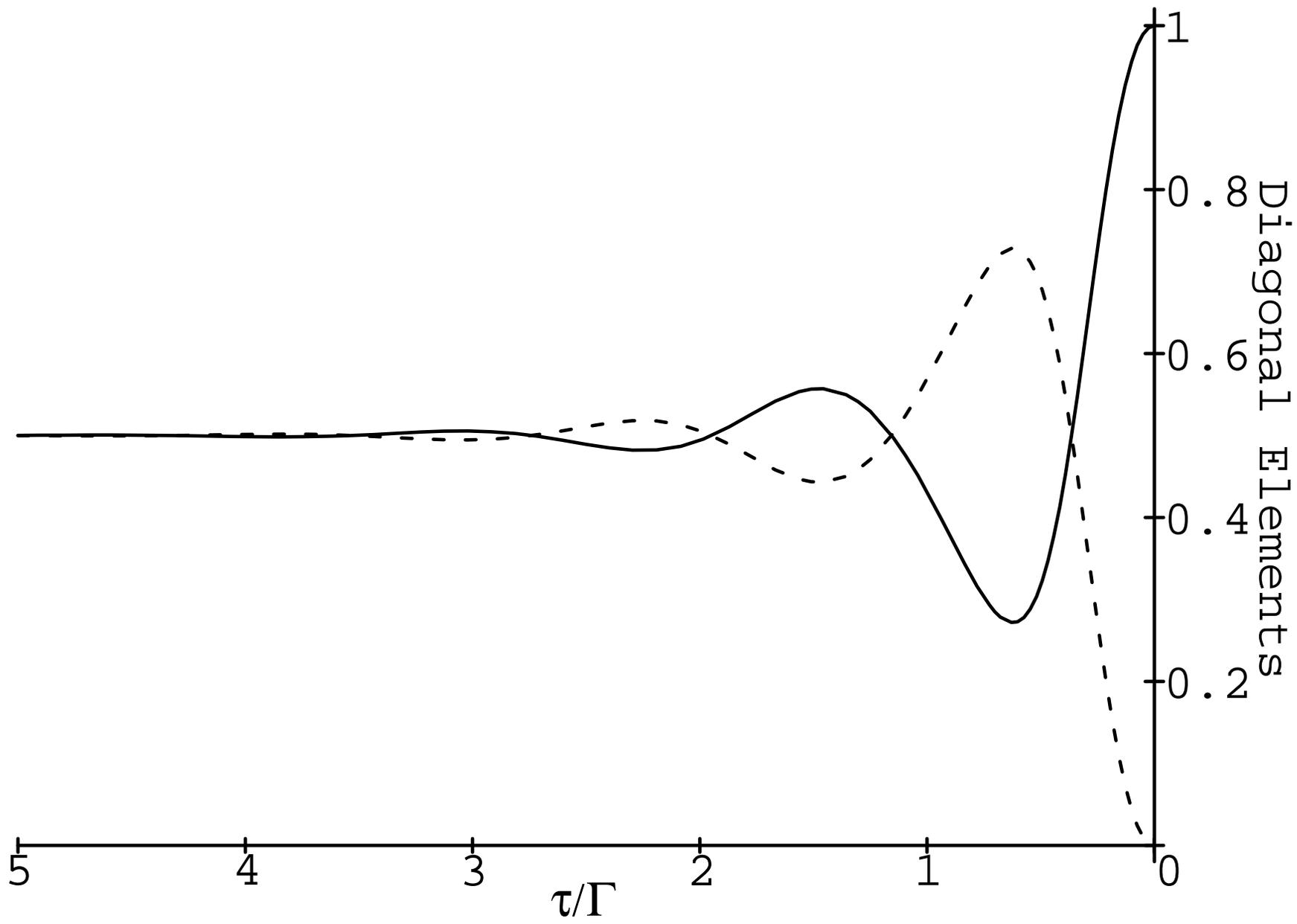

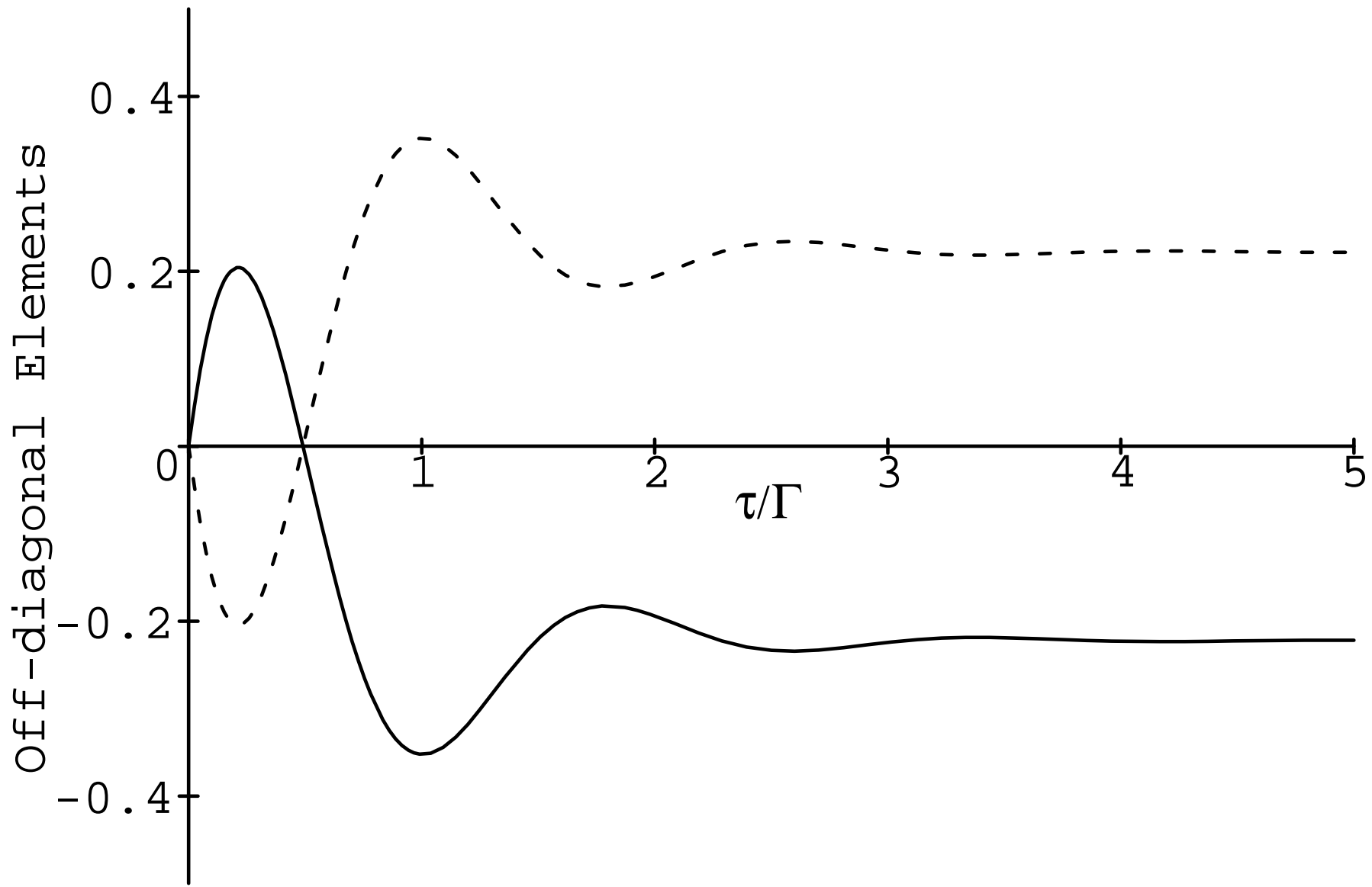

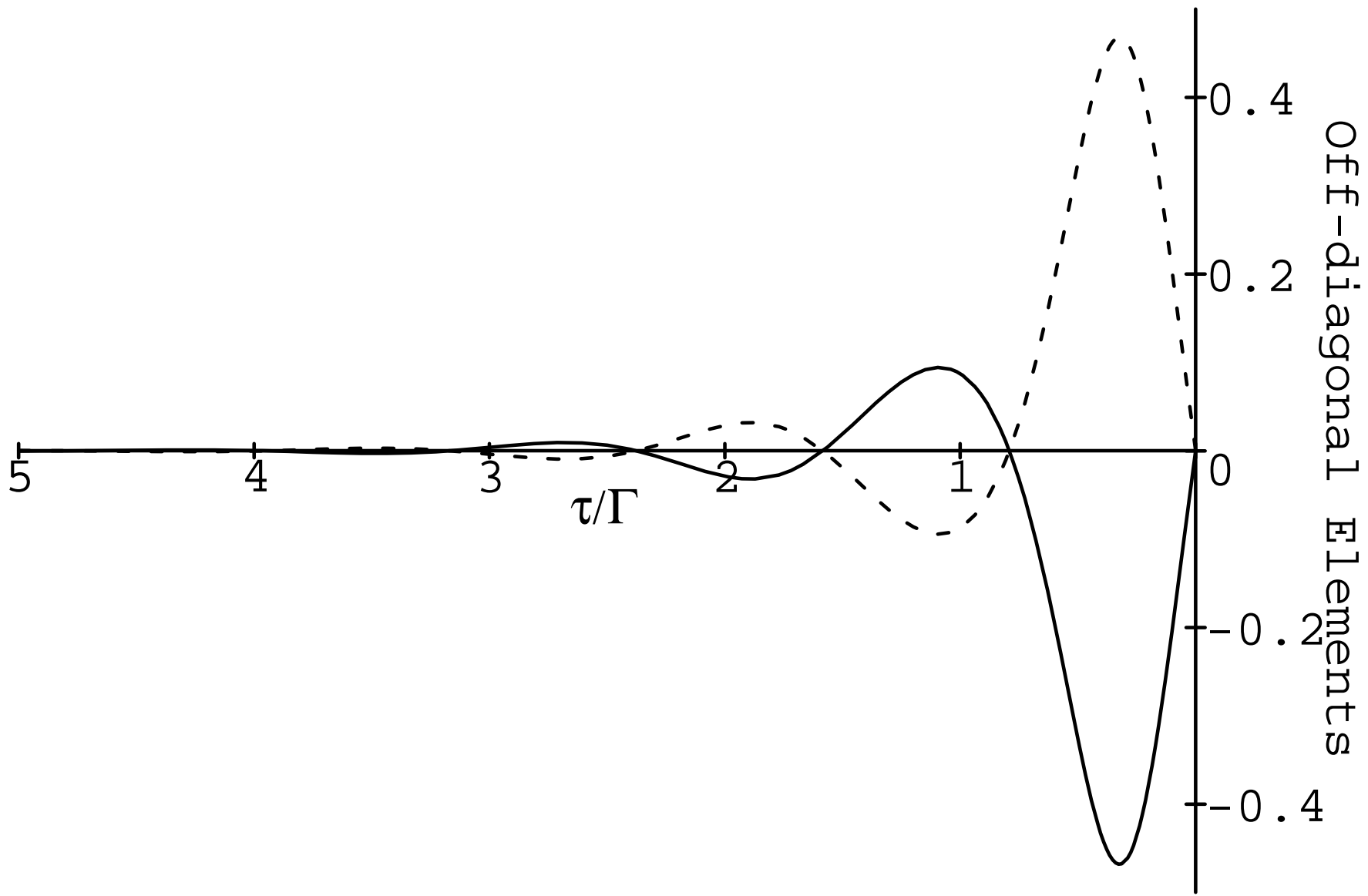

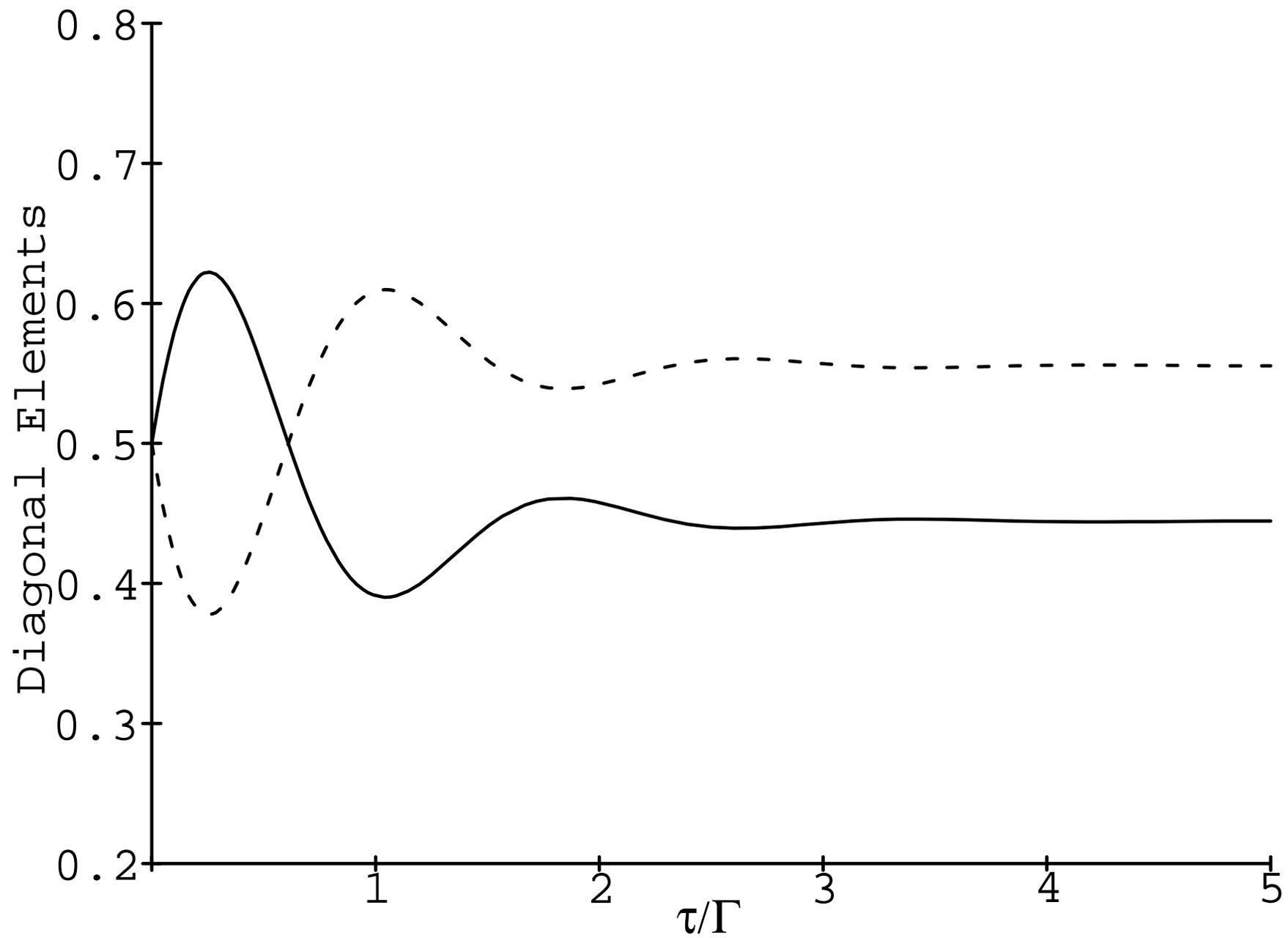

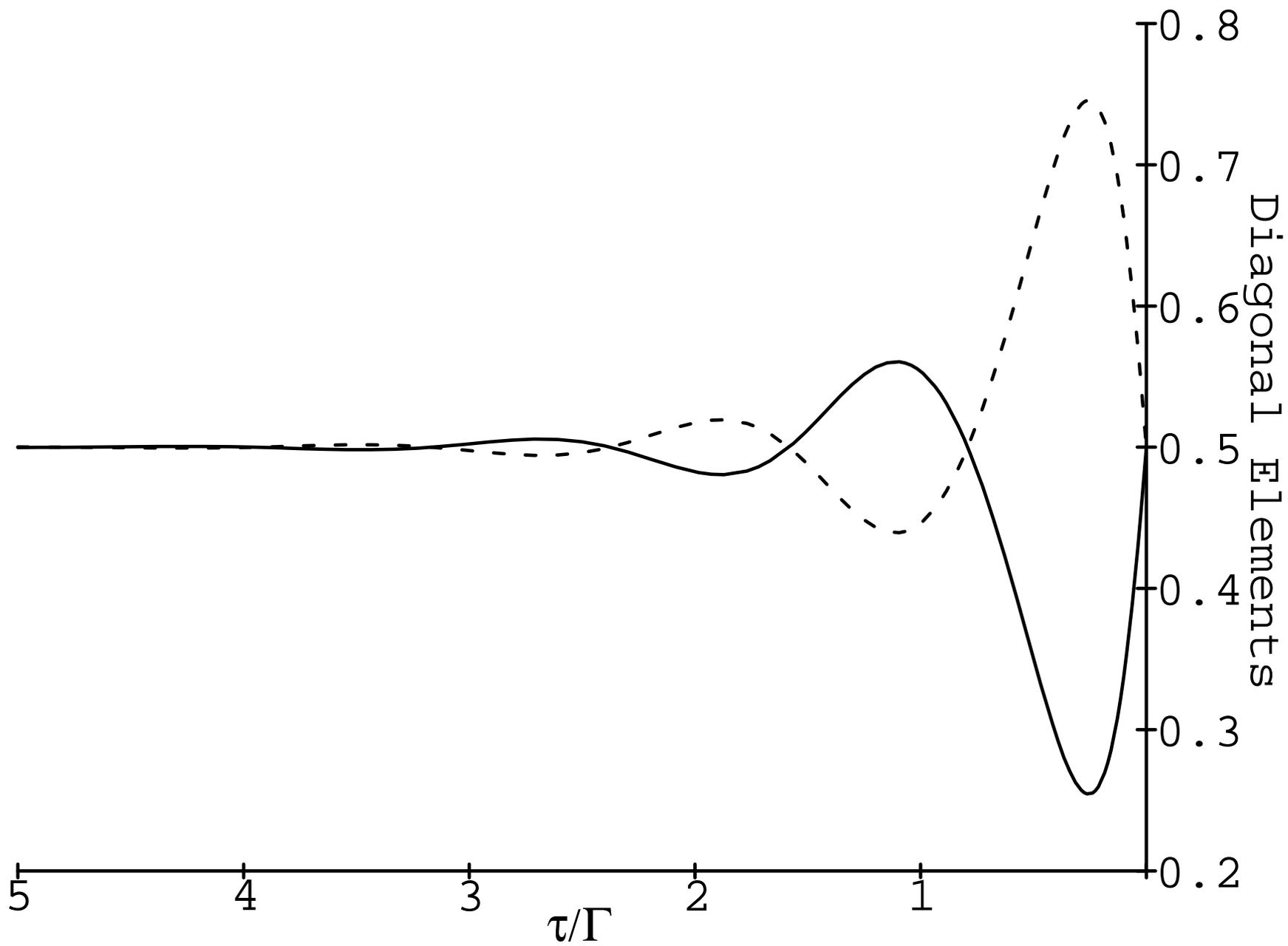

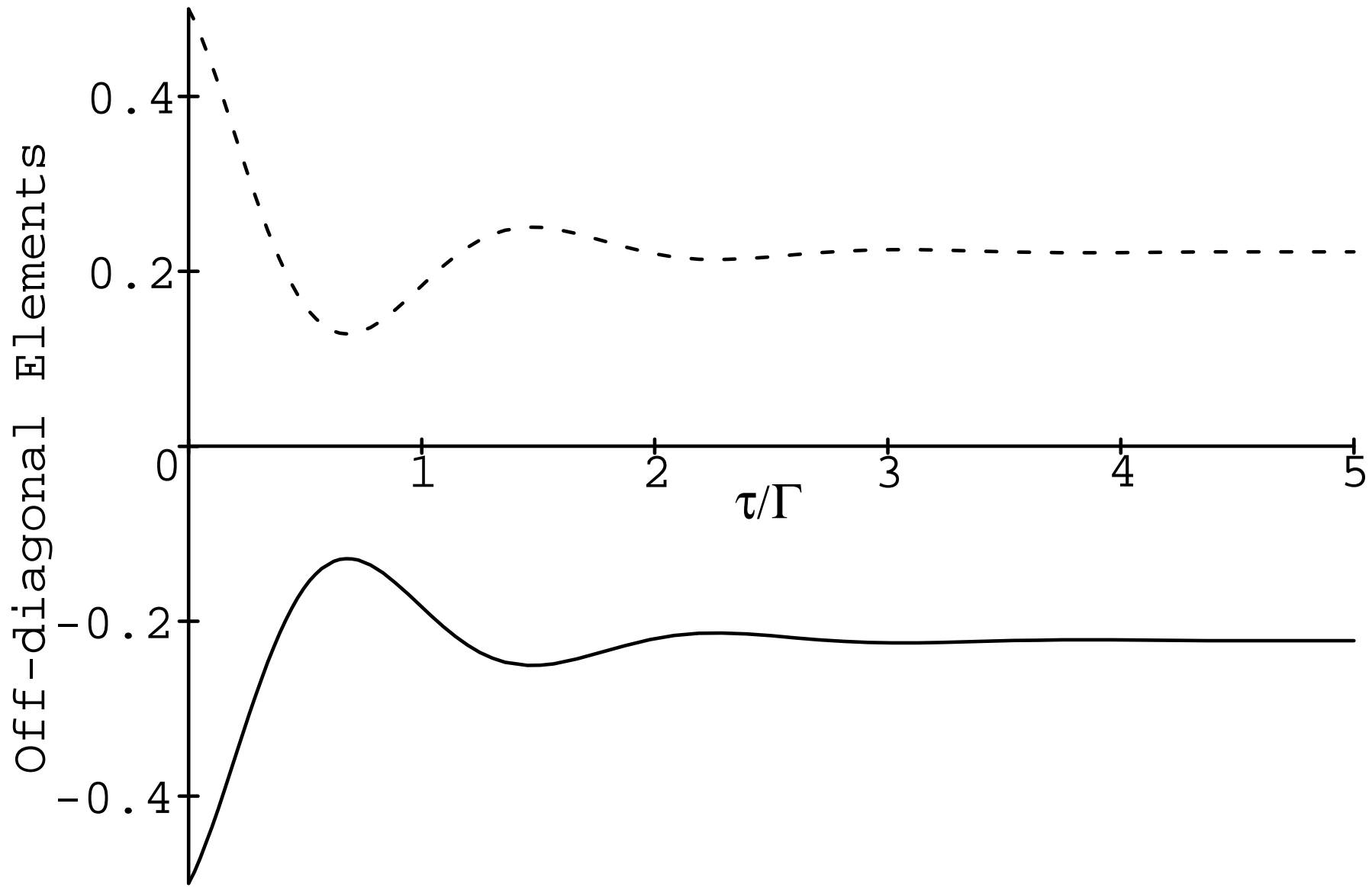

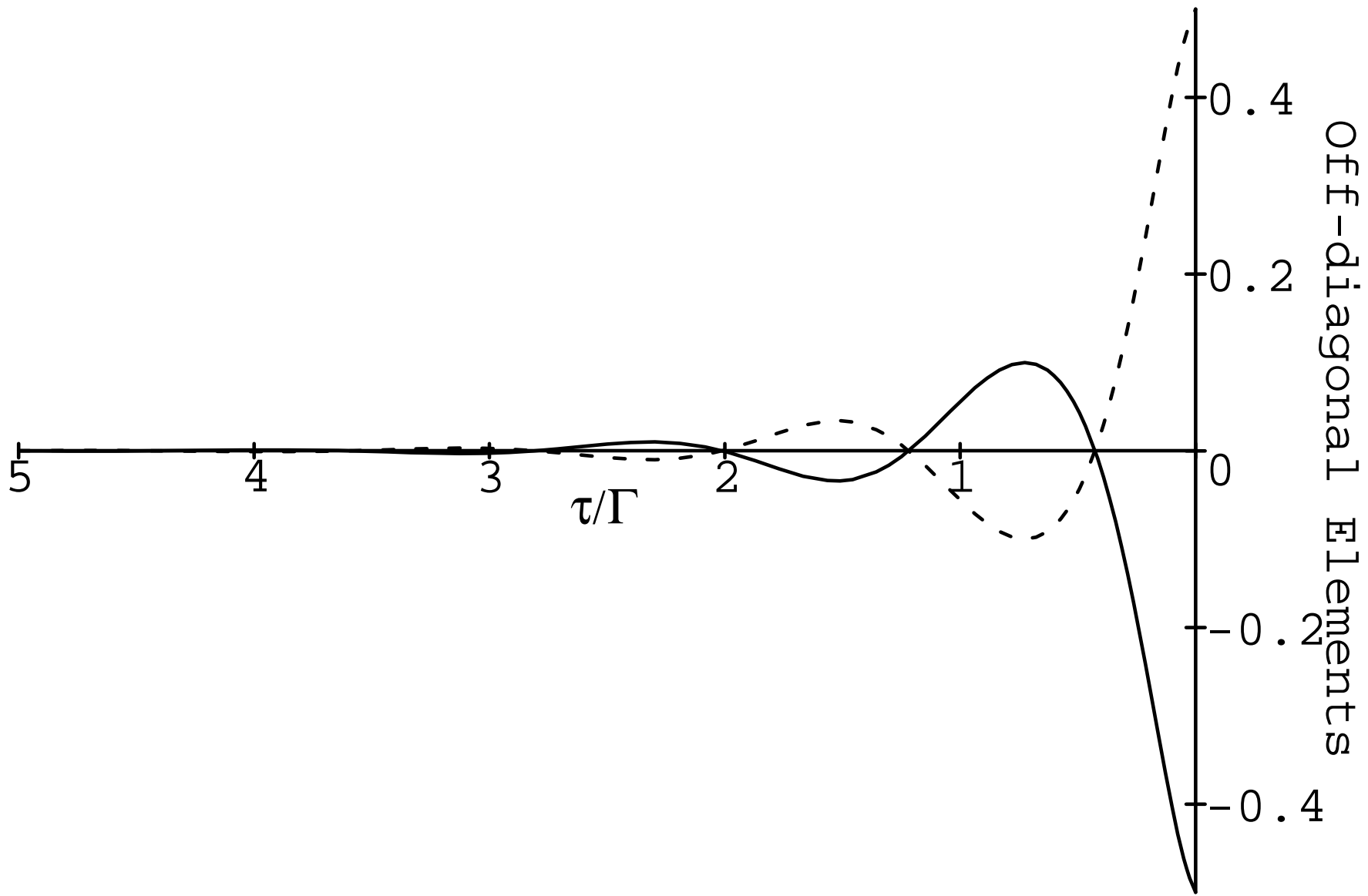

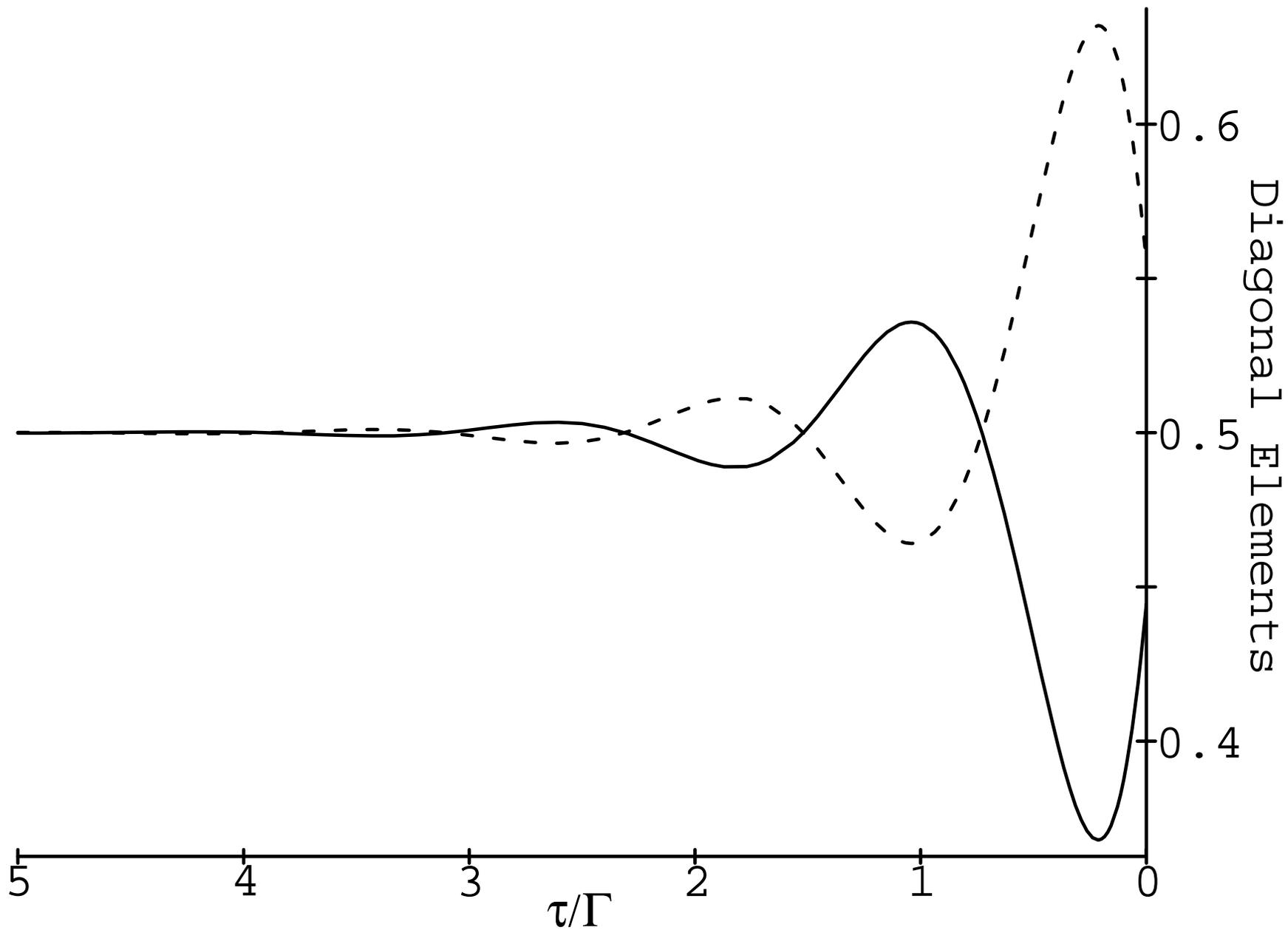

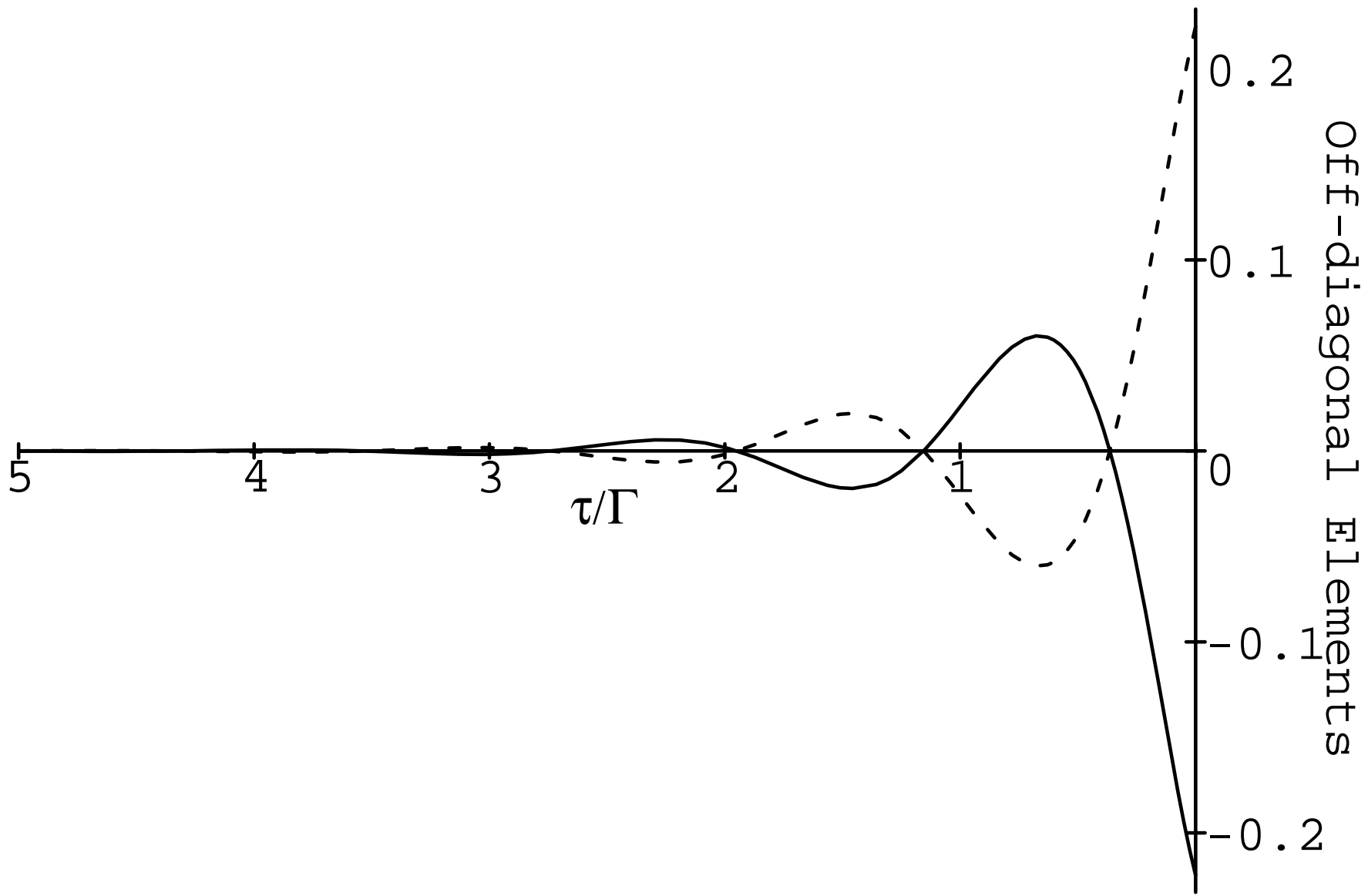